\documentclass{aa}  

\usepackage[normalem]{ulem}
\usepackage{natbib,twoopt}
\usepackage[breaklinks=true]{hyperref} 
\bibpunct{(}{)}{;}{a}{}{,} 
\makeatletter
\newcommandtwoopt{\citeads}[3][][]{\href{http://adsabs.harvard.edu/abs/#3}%
{\def\hyper@linkstart##1##2{}%
\let\hyper@linkend\@empty\citealp[#1][#2]{#3}}}
\newcommandtwoopt{\citepads}[3][][]{\href{http://adsabs.harvard.edu/abs/#3}%
{\def\hyper@linkstart##1##2{}%
\let\hyper@linkend\@empty\citep[#1][#2]{#3}}}
\newcommandtwoopt{\citetads}[3][][]{\href{http://adsabs.harvard.edu/abs/#3}%
{\def\hyper@linkstart##1##2{}%
\let\hyper@linkend\@empty\citet[#1][#2]{#3}}}
\newcommandtwoopt{\citeyearads}[3][][]%
{\href{http://adsabs.harvard.edu/abs/#3}
{\def\hyper@linkstart##1##2{}%
\let\hyper@linkend\@empty\citeyear[#1][#2]{#3}}}
\makeatother

\usepackage{soul}
\usepackage{graphicx}
\usepackage{txfonts}
\begin{document}

  \title{Spectral nuclear properties of NLS1 galaxies\thanks{The reduced spectra as FITS files are only available in electronic form at the CDS via anonymous ftp to cdsarc.u-strasbg.fr (130.79.128.5) or via http://cdsweb.u-strasbg.fr/cgi-bin/qcat?J/A+A/}}

   \author{E. O. Schmidt\inst{1}\fnmsep\thanks{Visiting Astronomer, Complejo 
Astron\'omico El Leoncito operated under agreement between the Consejo Nacional de Investigaciones Cient\'ificas y 
T\'ecnicas de la Rep\'ublica Argentina and the National Universities of La Plata, C\'ordoba, and San Juan.}, 
D. Ferreiro\inst{1,2}, L. Vega Neme\inst{1,2}, 
 and G. A. Oio\inst{1} }
          
  \institute{Instituto de Astronom\'ia Te\'orica y Experimental (IATE), Universidad Nacional de C\'ordoba,
 CONICET, Observatorio Astronómico de Córdoba, Laprida 854, Córdoba, Argentina \\
               \email{eduschmidt@oac.unc.edu.ar}
            \and            
            Observatorio Astron\'{o}mico, Universidad Nacional de C\'{o}rdoba,
Laprida 854, 5000 C\'{o}rdoba, Argentina 
                          }


 
 \abstract
   {It is not yet well known whether narrow line Seyfert 1 (NLS1) galaxies follow the $M_{BH} - \sigma_{\star}$ relation found for normal galaxies. Emission lines, such as [SII] or [OIII]$\lambda$5007, have been used as a surrogate of the stellar velocity dispersion and various results have been obtained. On the other hand, some active galactic nuclei (AGNs) have shown Balmer emission with an additional intermediate component (IC) besides the well-known narrow and broad lines. The properties of this IC have  not yet been fully studied.}
   {In order to re-examine the location of NLS1 in the $M_{BH} - \sigma_{\star}$ relation, we test some emission lines, such as the narrow component (NC) of H$\alpha$ and the forbidden [NII]$\lambda \lambda$6548,6584 and [SII]$\lambda \lambda$6716,6731 lines, as replacements for $\sigma_{\star}$. On the other hand, we study the properties of the IC of H$\alpha$ found in 14 galaxies of the sample to find a link between this component, the central engine, and the remaining lines. We also compare the emission among the broad component (BC) of H$\alpha$ and those emitted at the narrow line region (NLR) to detect differences in the ionizing source in each emitting region.}
   {We have obtained and studied medium-resolution spectra (170 km s$^{-1}$ FWHM at H$\alpha$) of 36 NLS1 galaxies in the optical range $\sim$5800 - 6800\AA . We performed a Gaussian decomposition of the H$\alpha +$[NII]$\lambda\lambda$6548,6584 profile to study the distinct components of H$\alpha$ and [NII] lines. We also measured the [SII] lines.}
   { We obtained black hole (BH) masses in the range log (M$_{BH}$/M$_{\odot}$) $= 5.4 - 7.5$ for our sample. We found that, in general, most of the galaxies lie below the $M_{BH} - \sigma_{\star}$ relation when the NC of H$\alpha$ and [NII] lines are used as a surrogate of $\sigma_{\star}$. The objects are closer to the relation when [SII] lines are used. Nevertheless, the galaxies are still below this relation and we do not see a clear correlation between the BH masses and FWHM$_{[SII]}$. Besides this, we found that 13 galaxies show an IC,  most of which are in the velocity range $\sim$ 700 $-$ 1500 km s$^{- 1}$. This is same range as in AGN types and is well correlated with the FWHM of BC and, therefore, with the BH mass. On the other hand, we found that there is a correlation between the luminosity of the BC of H$\alpha$ and NC, IC, [NII]$\lambda$6584, and [SII] lines.}
   {}


  \authorrunning{Schmidt et al.}\space\maketitle
%

\section{Introduction}
\label{Intro}

Narrow line Seyfert 1 galaxies are a subclass of active galactic nuclei (AGNs). The most common defining criterion for these objects is the width of the broad component of their optical Balmer emission lines in combination with the relative weakness of the [OIII]$\lambda$5007 emission with full width half maximum (FWHM) $\leq$ 2000 km s$^{-1}$ and the relative weakness of
[OIII]$\lambda$5007/H$\beta \leq$ 3. \citep{1985ApJ...297..166O, 1989ApJ...342..224G}. Optical spectroscopy classifications suggest that NLS1 represent about 15\% of the whole population of Seyfert 1 galaxies \citep{2002AJ....124.3042W}.

Observations of NLS1, as well as other AGNs, clearly show different types of emission lines in their spectra. Depending on the width of the emitted lines, they are classified as narrow lines with FWHM of a few hundred of km s$^{-1}$, or broad lines, with FWHM of a few thousand km s$^{-1}$. This implies two different regions of line formation, i.e, narrow line region (NLR) and the broad line region (BLR), respectively. 

In the middle of the '90s various works on QSOs suggested that the
traditional broad line region (BLR) consists of two components: one component with a FWHM of $\sim$  2000 km s$^{-1}$, called the
intermediate line region (ILR), and another very broad component
(VBLR) with a FWHM of $\sim$ 7000 km s$^{-1}$ 
blueshifted by $\geq$ 1000 km s$^{-1}$ \citep{1994ApJ...430..495B, 1994AJ....108.2016C}. 
Brotherton suggested that the intermediate line region (ILR) arises in a region inner to the
narrow line region (NLR). \cite{1996MNRAS.283L..26M} examined the profiles of H$\alpha$, H$\beta,$ and [OIII] lines and found evidence for an intermediate-velocity (FWHM $\leq$ 1000 km s$^{-1}$), line-emitting region that produces significant amounts of both permitted and forbidden lines. \cite{2007ApJ...659..250C} identified a line emission component with width FWHM = 1170 km s$^{-1}$ for the Sy1 NGC 4151, most probably originating between the BLR and NLR. \cite{2008ApJ...683L.115H} reported evidence for an intermediate line region in a sample of 568 quasar selected from the Sloan Digital Sky Survey (SDSS). They investigated the H$\beta$ and FeII emission lines and observed that the conventional broad H$\beta$ emission line 
could be decomposed into two components: one component with a very broad FWHM and the another with an intermediate FWHM. This ILR, whose kinematics seems to be dominated by infall, could be located in the outskirts of the BLR. In addition, \cite{2009ApJ...698..281C} detected ILR emission with width of 680 km s$^{-1}$ in the spectrum of NGC 5548. \cite{2010SCPMA..53.2307M} studied a sample of 211 narrow line Seyfert 1 galaxies selected from the SDSS, finding that the H$\beta$ profile can be fitted well by three (narrow, intermediate, and broad) Gaussian components with a ratio FWHM$_{BC}$ / FWHM$_{IC}$ $\sim$ 3 for the entire sample. Also, they suggested that the intermediate components originate from the inner edge of the torus, which is scaled by dust K-band reverberation. Related to this, there are some questions about the ILR. Is it linked to the central engine of the AGN? How is it related to the BLR emission? One of the aims of this paper is to answer these questions studying the intermediate component 
of H$\alpha$ in NLS1 galaxies.

On the other hand, and related to the central dynamics, according to several authors there is a correlation between the BH masses and the velocity dispersion of the stars of the bulge of the host galaxy \citep[e.g.,][]{2000ApJ...539L...9F,2000ApJ...539L..13G,2002ApJ...574..740T}. Because of the difficulty of measuring the stellar velocity dispersion in AGNs, sometimes emission lines are used as a surrogate. For example, in the case of the [OIII]$\lambda$5007 line, various authors found that NLS1 are off the $M_{BH} - \sigma_{\star}$ relation 
\citep[e.g.,][]{2001NewA....6..321M, 2004MNRAS.347..607B, 2004ApJ...606L..41G, 2005A&A...432..463M, 2005ApJ...633..688M, 2006ApJS..166..128Z} and on the contrary, other researchers found that these galaxies are on the relation \citep{2001A&A...377...52W, 2005MNRAS.356..789B}. \cite{2007ApJ...667L..33K} found that using [SII] lines as a replacement for the stellar velocity dispersion, NLS1 galaxies are on the relation. In this paper we re-examine the location of the NLS1 galaxies in the $M_{BH} - \sigma_{\star}$ relation using the narrow component of H$\alpha$ (originated in the NLR) and the forbidden emission lines [NII] and [SII] as surrogates of stellar velocity dispersion.    

As a whole, NLS1 galaxies represent a very interesting class of objects, and it is important to study
them both from their structural and dynamical point of view and
investigate how these objects fit into the unification schemes for AGNs. 
In order to study the intermediate component of H$\alpha$ and to test various emission lines as a replacement for $\sigma_{\star}$
we studied 36 NLS1 galaxies, of which 27 are in the Southern Hemisphere. We carried out a spectroscopic
study of these galaxies to analyze their internal kinematics and to
derive properties regarding their BLR, NLR, and ILR. In section \ref{observational} we present the sample and observations; a
description of how the
measurements were carried out is detailed in section \ref{measurements}; in section \ref{Dyn} we
present the results regarding the nuclear dynamics, for example,  the BH masses of the sample, location of the galaxies in the $M_{BH} - \sigma_{\star}$ relation, and the intermediate component of H$\alpha$,; and in section \ref{lumyfracpl} we show how the luminosities
of the different lines are related. We present some curious individual cases
in section \ref{curious.cases} and in section \ref{final} we draw our conclusions. Throughout this paper, we use 
the cosmological parameters $H_{0} = 70$ Km s$^{-1}$, $\Omega_{M}= 0.3$, and $\Omega_{\Lambda}= 0.7$.

\section{Observational data}
\label{observational}

\subsection{Sample and observations}
\label{Sample}

We have selected NLS1 galaxies from the V\'eron \&
V\'eron catalog \citep{2010A&A...518A..10V} with redshifts z$<$0.15 and brighter than
$m_{b}$ $<$ 18 with $\delta \le 10^{\circ}$.  Of those objects, we chose galaxies whose nuclear kinematics are poorly
or even not studied. This way our sample consists of 
36 galaxies, of which 27 are in the Southern Hemisphere, while the remaining have $0^{\circ}$ $<$ $\delta \le
10^{\circ}$. This coordinate range allows us to observe the sample from the Complejo AStronómico el LEOncito (CASLEO), in Argentina.

Observations were performed in different campaigns between 2011
and 2014. We carried out long-slit spectroscopy using the REOSC
spectrograph at CASLEO 2.15 m. Ritchey-Chr\'etien telescope, San Juan,
Argentina. The spectrograph has  a Tektronix 1024 $\times$
1024 CCD attached with 24 $\mu$m pixels. The galaxies were observed using a 2.7
arsec wide slit and the extractions of each spectrum were of $\sim$
2.3 arsec. For a mean distance of 240 Mpc (mean redshift of
$\sim$0.06) the spectra correspond to the central $\sim$3 kpc in
projected distance. We used a 600 line mm$^{-1}$ grating giving a resolution
of 170 km \ s$^{-1}$ FWHM around H$\alpha$. The observations cover the spectral
range from 5800\AA \ to 6800\AA. 
The spectra were calibrated in wavelengths using comparison
lines from a Cu-Ne-Ar lamp and three standard stars were observed each
night to flux calibrate the spectra. Standard data reduction
techniques were used to process the data, mainly {\it ccdred} and {\it
  longslit} packages included in IRAF\footnote{IRAF: the Image
  Reduction and Analysis Facility is distributed by the National
  Optical Astronomy Observatories, which is operated by the
  Association of Universities for Research in Astronomy, Inc. (AURA)
  under cooperative agreement with the National Science Foundation
  (NSF).}. The obtained spectra have a mean signal to noise of $\sim$ 16 around 6000 \AA. Although some galaxies in our sample already have spectra, in most cases
  they have a lower resolution or are not flux calibrated
  (as in the cases of 2df and 6df spectra). In our sample, 11 objects have SDSS spectra
  with a very similar resolution. They were taken with the SDSS spectrograph, which has a fiber diameter of 3 arcsec.
  Table \ref{tabla1} lists the galaxy name, right ascension,
declination, apparent B magnitude, redshift, date of the observation, and exposure time.

\begin{table*}
\caption{List of observed galaxies. Column 1: galaxy name. Columns 2 to 5: right ascension (J2000),
  declination (J2000), apparent B magnitude, and redshift, as taken
  from NED. Columns 6 and 7: observation date and exposure time.}
\begin{center}
\begin{tabular}[h]{|l|c|c|c|c|c|c|}
\hline
\hline
                         & R.A          & Decl.    &            &                        &           & Exposure Time              \\
Galaxy                  & J(2000)       & J(2000)  &  $m_{b}$  & z                       & Date      & (sec)                      \\  \hline
 
1RXS J040443.5$-$295316    & 04 04 43.1 &       $-$29 53 23  &  17.3   &        0.06002         & 08$-$09$-$2013  &       3$\times$2400\\

2MASXJ01115114$-$4045426   & 01 11 51.1 &       $-$40 45 43  &  16.9   &        0.05429         & 11$-$01$-$2011  &       3$\times$1800\\
2MASX J01413249$-$1528016  & 01 41 32.5 &       $-$15 28 01  &  16.9   &        0.08106         & 10$-$27$-$2014        &       3$\times$1800\\

2MASXJ05014863$-$2253232   & 05 01 48.6 &       $-$22 53 23  &  14.0   &        0.04080         & 10$-$29$-$2011  &       3$\times$1800\\
2MASX J21124490$-$3730119  & 21 12 44.9 &       $-$37 30 12  &  17.0   &        0.04297         & 08$-$17$-$2012        &       3$\times$1800\\

2MASX J21531910$-$1514111  & 21 53 19.1 &       $-$15 14 12  &  14.7   &        0.07780         & 10$-$25$-$2014  &       2$\times$2400\\
2MASXJ21565663$-$1139314   & 21 56 56.5 &       $-$11 39 32  &  15.4   &        0.02808         & 09$-$07$-$2013  &       3$\times$2400\\

6dF J1117042$-$290233      & 11 17 04.2 &       $-$29 02 33  &  15.8   &       0.07046           & 04$-$26$-$2013        &       4$\times$1800\\
CTSJ13.12                & 13 51 29.5 & $-$18 13 47  &  17.5   &        0.01221         & 04$-$23$-$2012  &       3$\times$1800\\

CTSM02.47                & 10 46 23.5 & $-$30 04 20  &  17.0   &        0.05706         & 04$-$25$-$2012  &       3$\times$1800\\
EUVEJ0349$-$53.7           & 03 49 28.5 &       $-$53 44 47  &  15.7   &        0.13000         & 10$-$27$-$2014  &       3$\times$2400\\

FAIRALL0107               & 21 35 29.5 &        $-$62 30 07  &  16.7   &        0.06096         & 01$-$11$-$2011  &       3$\times$1800\\
HE1107$+$0129            & 11 10 12.1 &         $+$01 13 27  &  16.5   &        0.09550         & 09$-$04$-$2013  &       2$\times$2400\\

HE1438$-$0159            & 14 41 11.5 &         $-$02 12 35  &  16.6   &        0.08292         & 02$-$05$-$2014  &       3$\times$2400\\
IGRJ16185$-$5928         & 16 18 36.4 &         $-$59 27 18  &  16.5   &        0.03462         & 04$-$25$-$2012  &       3$\times$1800\\

IRAS04576$+$0912        & 05 00 20.8 &  $+$09 16 56  &  16.6   &        0.03609         & 01$-$11$-$2011        &       2$\times$1800\\
IRAS16355$-$2049         & 16 38 30.9 &         $-$20 55 25  &  14.5   &        0.02637         & 08$-$15$-$2012        &       3$\times$1800\\

MCG$-$04.24.017          & 10 55 55.4 &         $-$23 03 25  &  14.5   &        0.01281         & 04$-$24$-$2012  &       3$\times$2400\\
MCG$-$05.01.013          & 23 53 27.9 &         $-$30 27 40  &  15.8   &        0.03068         & 09$-$09$-$2013  &       3$\times$2400\\

NPM1G$-$17.0312            & 11 40 42.2 &       $-$17 40 10  &  15.9   &        0.02187         & 04$-$23$-$2012        &       4$\times$1800\\
RBS1665                  & 20 00 15.5 & $-$54 17 12  &  16.5   &        0.06070         & 04$-$25$-$2012  &       3$\times$1800\\

RX J0024.7$+$0820          & 00 24 45.7 &       $+$08 20 57  &  17.2   &        0.06700         & 10$-$25$-$2014  &       3$\times$2400\\
RX J0323.2$-$4931          & 03 23 15.3 &       $-$49 31 07  &  16.5   &        0.07100          & 10$-$29$-$2011        &       3$\times$1800\\

RX J0902.5$-$0700          & 09 02 33.6 &       $-$07 00 04  &  17.7   &         0.08911         & 04$-$09$-$2013        &       2$\times$2400\\
RX J2301.8$-$5508          & 23 01 52.0 &       $-$55 08 31  &  14.7   &        0.14100         & 10$-$27$-$2014  &       3$\times$2400\\

SDSS J134524.69$-$025939.8 & 13 45 24.7 &     $-$02 59 40   &   17.1   &         0.08552         & 04$-$09$-$2013        &       3$\times$2400\\

SDSS J144052.60$-$023506.2  & 14 40 52.6 &    $-$02 35 06  &    16.3   &        0.04434         & 04$-$30$-$2014        &       3$\times$2400\\
SDSS J151024.92$+$005843.9  & 15 10 24.9 &    $+$00 58 44  &    17.3   &        0.07215         & 04$-$06$-$2013  &       2$\times$1800\\

SDSS J153001.82$-$020415.1  & 15 30 01.8 &    $-$02 04 15   &   17.1   &        0.05118         & 05$-$02$-$2014 &       3$\times$2400\\
SDSS J153705.95$+$005522.8 & 15 37 05.9  &    $+$00 55 23  &    17.3   &        0.13655         & 04$-$09$-$2013        &       3$\times$2400\\

SDSS J225452.22$+$004631.4 & 22 54 52.2  &    $+$00 46 31  &    17.4   &        0.09073          & 08$-$09$-$2013        &       3$\times$2400\\
V961349$-$439            & 13 52 59.6 &        $-$44 13 25  &   15.4   &        0.05200         & 04$-$06$-$2013  &       2$\times$1800\\

WPV8507                 & 00 39 15.8 &        $-$51 17 01  &    15.8   &        0.02861         & 08$-$18$-$2012  &       3$\times$1800\\
Zw037.022                & 10 23 59.8 &        $+$06 29 07  &   16.0   &        0.04385         & 04$-$30$-$2014        &       3$\times$2400\\
Zw049.106                & 15 17 51.7 &        $+$05 06 28  &   17.5   &        0.03878         & 04$-$30$-$2014  &       3$\times$2400\\
Zw374.029                & 20 55 22.3 &        $+$02 21 16  &   15.1   &        0.01356         & 04$-$23$-$2012        &       2$\times$1800\\
\hline
\end{tabular}

\label{tabla1}
\end{center}
\end{table*}

\

\subsection{NLS1 spectra}
\label{Spectra}

Here we present the spectra of 36 NLS1s for the red spectral region
5800\AA \ - 6800\AA. In this work we focus on this spectral region to study
in detail the emission lines such as H$\alpha+$[NII]$\lambda\lambda$6548,6584 and
[SII]$\lambda\lambda$6716,6731.

The beginning of the spectral region
presented conforms to the higher initial wavelength for the red
observations, while there were usually sky absorption
lines near 7000\AA. This made the measurement of sulfur lines difficult and, depending
on the
redshift, in many cases prevented these measurements, thus making 5800\AA\ -
6800\AA\ a compromise between these constrains.

\begin{figure*}
  \includegraphics[trim = 50mm 10mm 00mm 00mm, clip, width=1\linewidth]{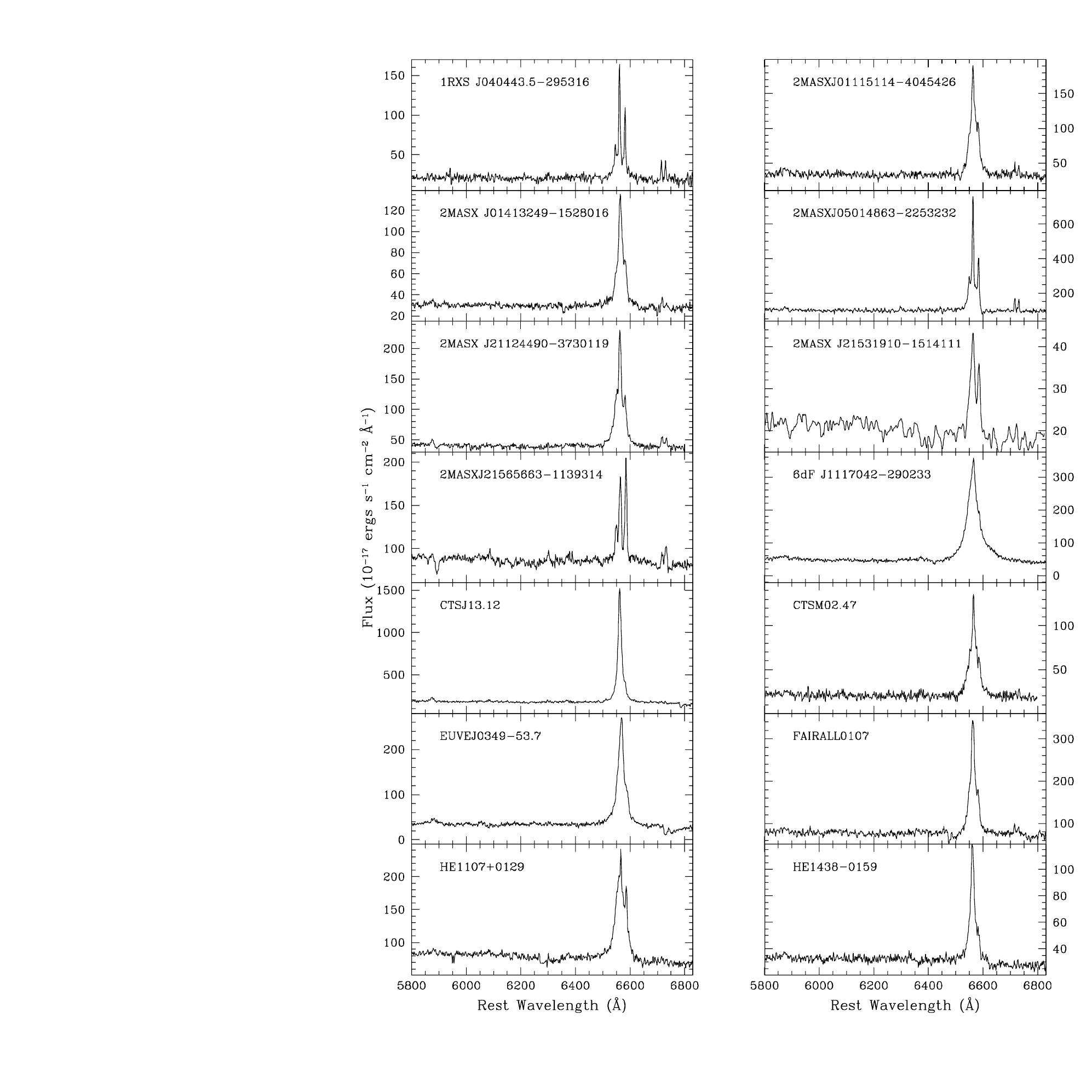}
 \caption{Observed spectra of the NLS1 in the range 5800 \AA{} $-$ 6800 \AA{} at rest frame.}
 \label{fig:espectros_A}
\end{figure*}

\begin{figure*}
  \includegraphics[trim = 50mm 10mm 0mm 00mm, clip, width=1\textwidth]{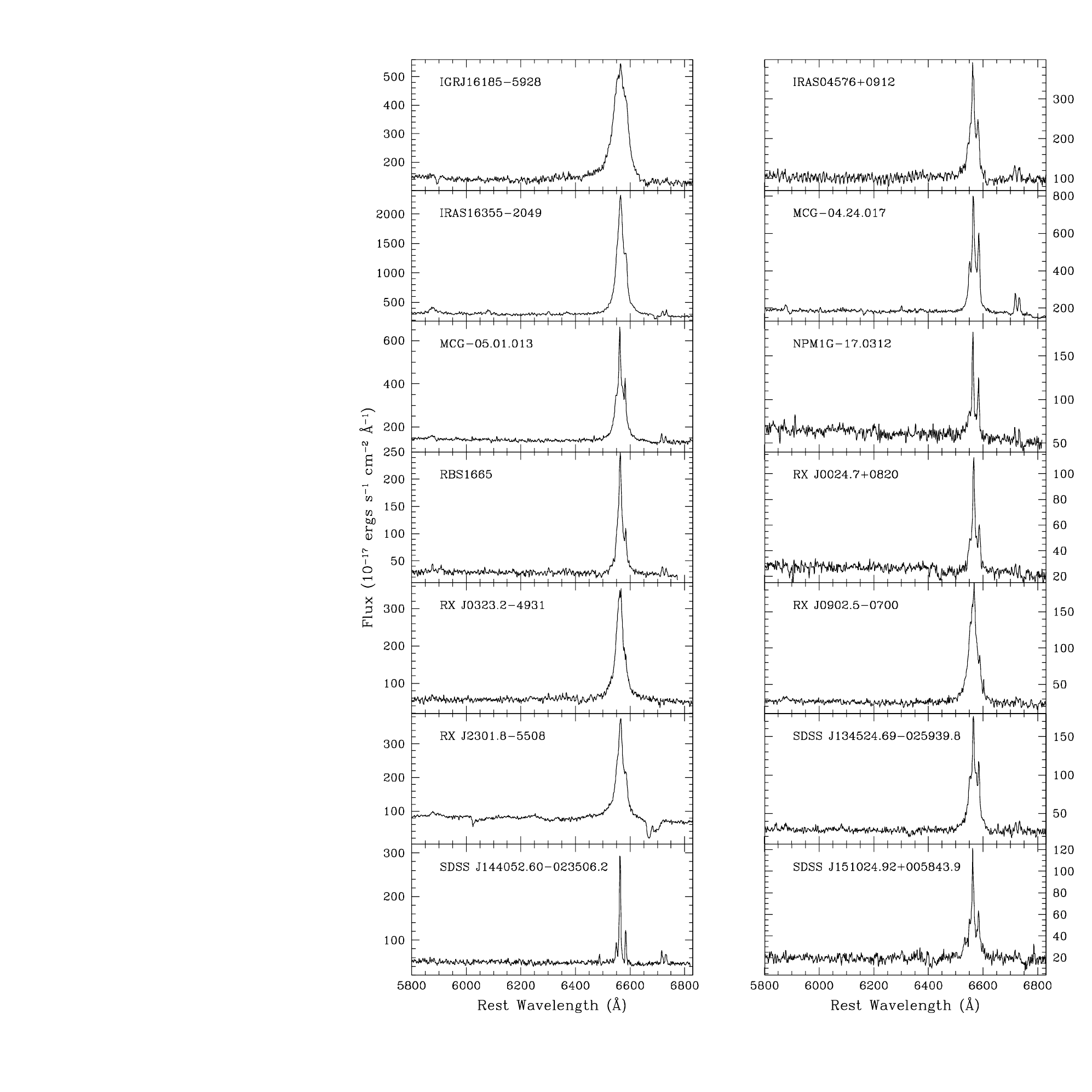}
 \caption{Observed spectra of the NLS1 in the range 5800 \AA{} $-$ 6800 \AA{} at rest frame.}
 \label{fig:espectros_B}
\end{figure*}

\begin{figure*}

  \includegraphics[trim = 50mm 80mm 0mm 00mm, clip, width=1\textwidth]{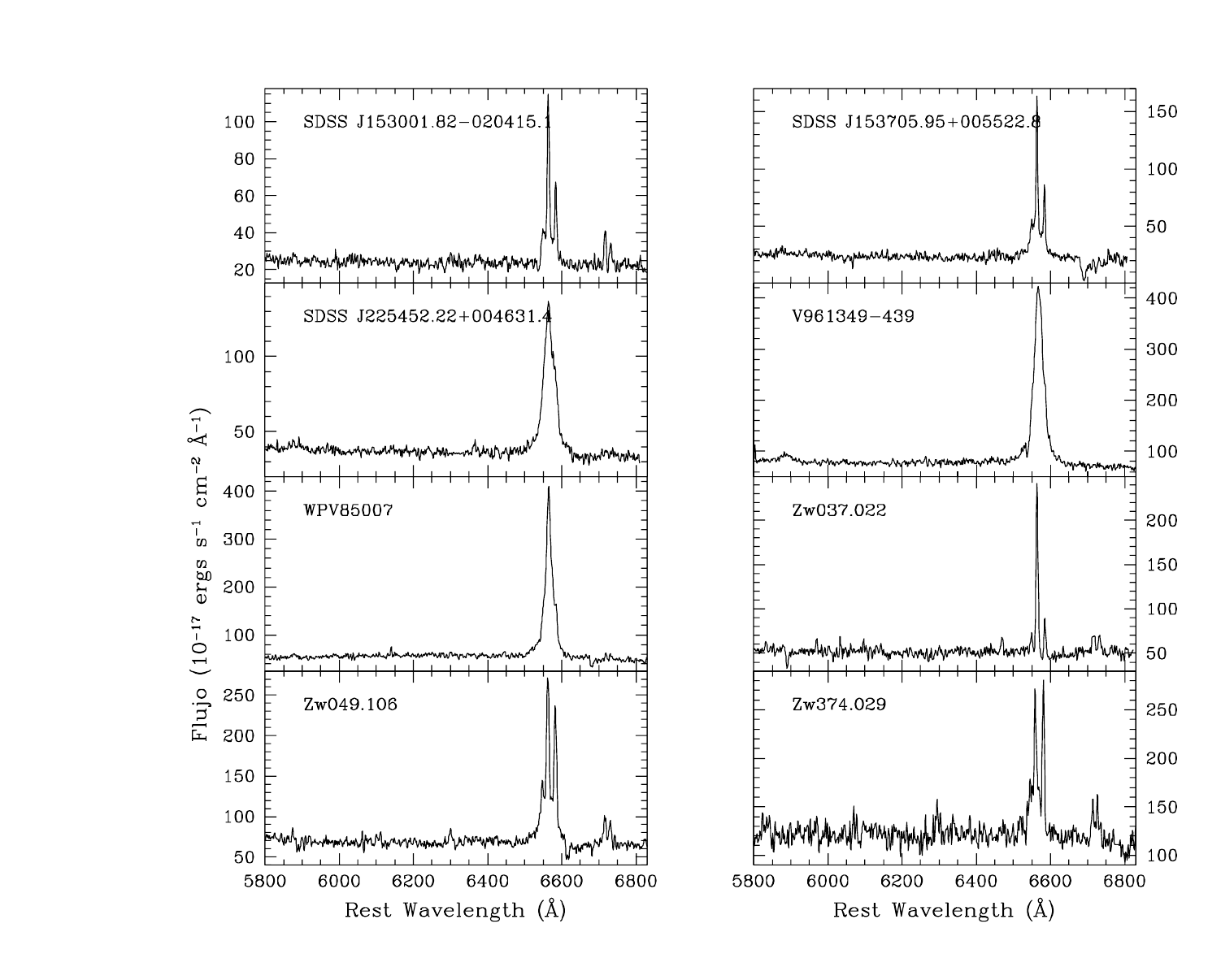}
 \caption{Observed spectra of the NLS1 in the range 5800 \AA{} $-$ 6800 \AA{} at rest frame.}
 \label{fig:espectros_C}
\end{figure*}

Figures \ref{fig:espectros_A}, \ref{fig:espectros_B}, and
\ref{fig:espectros_C} present the spectra of the NLS1s for the red
spectral region. All spectra are in rest frame for which we used the
published redshift data from the NASA/IPAC Extragalactic Database
(NED), and we present their fluxes in units of 10$^{-17}$ ergs\ cm$^{-2}$
s$^{-1}$ \AA$^{-1}$. As expected from photoionized gas in active nuclei,
H$\alpha+$[NII] emission lines are the most important feature in all
the objects, while in 25 out of 36 spectra
[SII]$\lambda\lambda$6716,6731 lines are easily detected. We also
identify [OI]$\lambda$6300 (over 3$\sigma$) in only seven objects. Sodium
lines Na$\lambda\lambda$5890, 5896 were detected, although these (as well
as other absorption lines) are often weak mainly because of the dilution
of the stellar features by a nonstellar continuum; these lines also appear in emission in some cases.

\section{Emission line measurements}
\label{measurements}

Our interest resides in the emission lines, which offer
important information about kinematics and luminosities of the central
regions of NLS1s. Emission line profiles of NLS1s can be represented by a single or a combination
of Gaussian profiles. We used the LINER routine
\citep{1993Pogge..and..Owen}), which is a $\chi^2$ minimization
algorithm that can fit several Gaussians to a line profile. We adopted
a procedure of fitting three possible Gaussians to H$\alpha$: one
broad component (BC) to the wings of the line, usually extending to
$\sim$5000 km s$^{-1}$ in the base and with FWHM around 2500
km s$^{-1}$; one narrow component (NC) fitted to the peak of the
profile with typically FWHM$\sim$300 km s$^{-1}$; and an intermediate
component (IC) with FWHM between those of the NC and BC (IC is
explained in detail in Sect. \ref{intermediate}). It was not necessary to apply any Gaussian decomposition for the
[NII]$\lambda\lambda$6548,6584 and [SII]$\lambda\lambda$6716,6731
lines, since one
component for each line profile fits well. Two
constraints were used for [NII] lines. First, [NII]$\lambda$6548 and
[NII]$\lambda$6584 should be of similar FWHM because both lines are emitted
in the same region and, second, the flux ratio should be equal
to their theoretical value 1:3. Similar FWHM constraints as for [NII] were applied for [SII] lines.
By this procedure, we obtained FWHM, center, and peak of each line, fully describing the Gaussians of
H$\alpha$, [NII], and [SII] lines. Typical uncertainties of the FWHM measurements are $\sim$ 10 \%, while for the fluxes these are on the order
of $\sim$ 15 \%.

\begin{figure*}
 \begin{minipage}{\linewidth}
 \begin{center}
  \includegraphics[trim = 3mm 12mm 3mm 130mm, clip, width=1\textwidth]{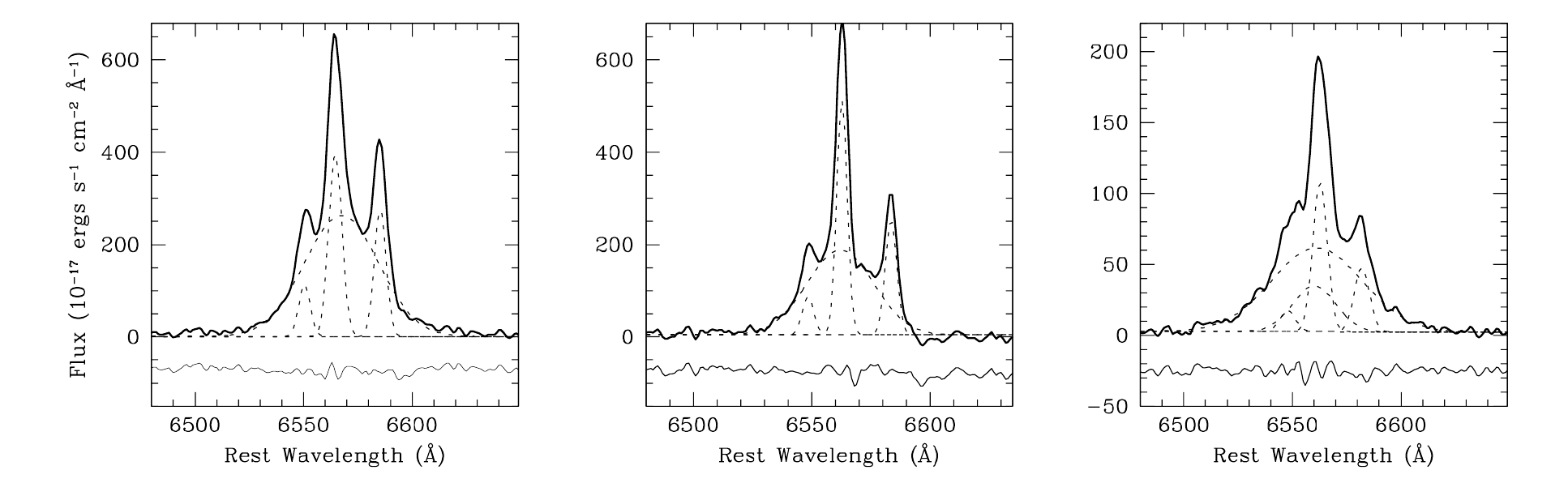}
  \end{center}
 \caption{ From left to right: Gaussian decomposition for the galaxies MCG04.24.017,
   2MASXJ05014863$-$2253232, and 2MASX J21124490$-$3730119, respectively. Thick line represents the spectra; the H${\alpha}$ components and [NII]$\lambda$6548,6584 are shown in dotted lines. The residuals are plotted at the bottom of each panel and were displaced from zero for clarity.}
   
\label{fig:Ej_liner}
 \end{minipage}
\end{figure*}

\begin{figure}
 \begin{minipage}{\linewidth}
 \begin{center}
 \includegraphics[trim = 0mm 00mm 0mm 00mm, clip, width=1\textwidth]{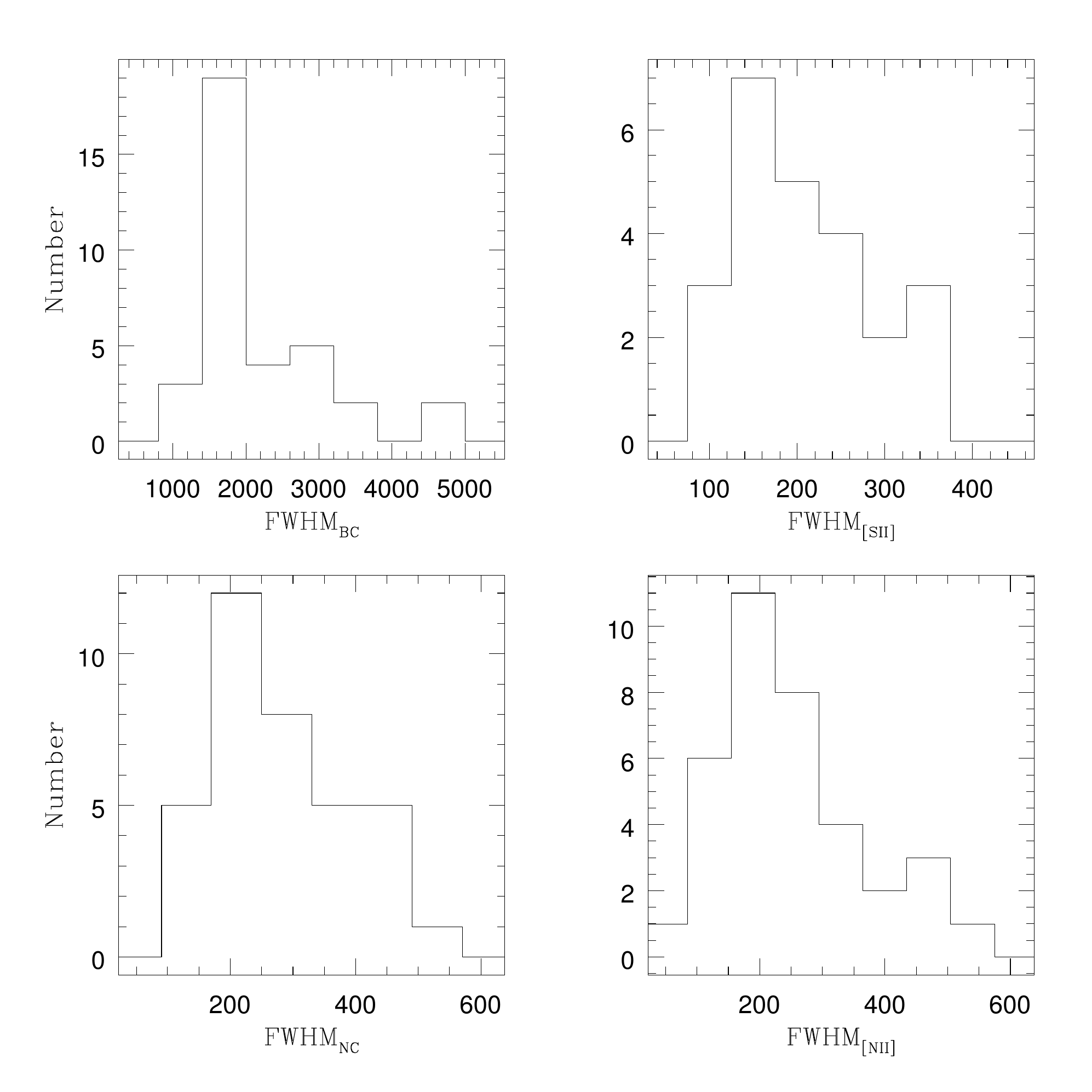}
  \end{center}
 \caption{Histograms of the FWHM of the emission lines in km  s$^{-1}$. From left to right and top to bottom: FWHM of BC, [SII], NC and [NII]. All FWHM were corrected by the instrumental width.}
\label{fig:Hist.fwhm}
 \end{minipage}
\end{figure}

Figure \ref{fig:Ej_liner} shows three typical fits for H$\alpha +$
[NII]$\lambda$6548, 6584. Two of these are the same as the two components
for H$\alpha$ and one fit has the IC line. The solid thick line
represents the spectra and the dotted lines represent each Gaussian profile
fitted. The thin line represents the residuals of the fit (plotted below), which
are similar to the noise level around the line. Figure
\ref{fig:Hist.fwhm} shows the distributions of the FWHM of
H$\alpha$ BC, H$\alpha$ NC, [NII], and
[SII]. All FWHM were corrected by the instrumental width as FWHM$^{2}$ $=$ FWHM$_{obs}^{2}$ -
FWHM$_{inst}^{2}$, where FWHM$_{obs}$ is the measured FWHM and FWHM$_{inst}$ is the instrumental broadening
($\sim$3.8\AA\ or 170 km s$^{-1}$).

\section{Nuclear dynamics}
\label{Dyn}

\subsection{Virial BH masses}
\label{virial}

Black holes are very important in the study of active galaxies. They
constitute a fundamental parameter to understand the mechanisms
involved in the nuclear regions of AGNs and could also provide information about the process of galaxy formation. If we assume that
the BLR clouds are isotropically spatially and kinematically
distributed, $M_{BH}$ is related to the radius of the BLR ($R_{BLR}$) and the mean cloud velocity $v$ inside it 
as $M_{BH}=R_{BLR}\ v^{2}\ G^{-1}$.
According to \cite{2005ApJ...630..122G}, BH masses can be calculated through
luminosity and FWHM of the same Balmer line, as follows:

\begin{equation}
\label{eq:GH}
M_{BH} =  2 \times 10^{6} \left(\frac{L_{H \alpha}}{10^{42} ergs \ s^{-1}}\right)^{0.55}  \left(\frac{FWHM_{H \alpha}}{10^{3} km \ s^{-1}}\right)^{2.06}  M_{\odot}
.\end{equation}

\

\begin{figure}
 \begin{minipage}{\linewidth}
 \begin{center}
  \includegraphics[trim = 00mm 0mm 0mm 0mm, clip, width=1\textwidth]{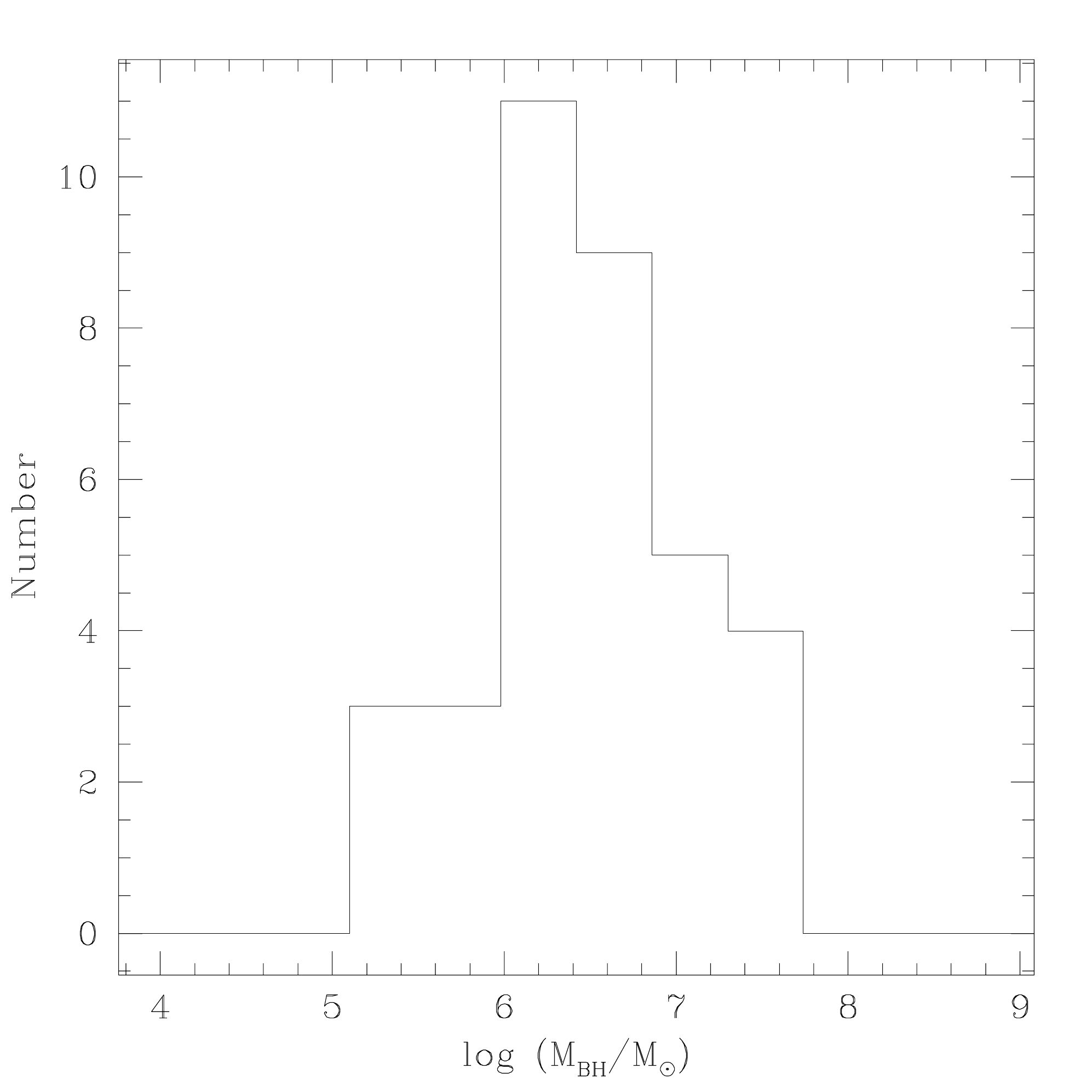}
  \end{center}
  \caption{Histogram of the black hole masses of the sample.}
 \label{fig:Hist.acrecion}
 \end{minipage}
\end{figure}

\noindent This method is highly useful because Balmer lines are easily
detectable even in distant AGNs. However, this equation can be applied
only to AGNs that show broad component lines, thereby the importance
of performing Gaussian deconvolution to Balmer lines such as described in section \ref{measurements}. Using equation \ref{eq:GH} we
determined M$_{BH}$ of our sample by taking into account the BC of
H$\alpha$ after correcting for instrumental resolution. Table \ref{tabla.masas} specifies the galaxy name, FWHM, and luminosity of BC and
BH masses. Figure \ref{fig:Hist.acrecion} shows the distribution
of BH masses, whose typical uncertainties are around $\sim$ 0.1 dex. Almost all of the BH masses are between
log (M$_{BH}$/M$_{\odot}$) $\sim 5.75$ and $7.35$ (mean value $\sim
6.2$). Only two objects show masses log (M$_{BH}$/M$_{\odot}$) $\geq 7.5$; these objects are 6dF J1117042$-$290233, RX J2301.8$-$5508, which have log (M$_{BH}$/M$_{\odot}$) of 7.7 and 7.5, respectively. There are also a few determinations of low-mass
BHs with log (M$_{BH}$/M$_{\odot}$) $<$ 5.5, mainly due to their low BC
luminosities. These galaxies are SDSS J144052.60$-$023506.2, Zw374.029 with log (M$_{BH}$/M$_{\odot}$) $=$ 5.4 and 
NPM1G$-$17.0312 with log (M$_{BH}$/M$_{\odot}$) $=$ 5.3.

\begin{table*}
\caption{FWHM of the BC of H${\alpha}$, luminosity of that component, and BH masses. The galaxy Zw037.022 does not show the broad component (see Section \ref{curious.cases})}
\begin{center}
\begin{tabular}[h]{|l|l|l|l|l}
\hline
\hline
                         & FWHM$_{BC}$       & log L$_{BC}$  &   log M$_{BH}$                  \\
Galaxy                  & (km  s$^{-1}$ )       & (ergs  s$^{-1}$)  &  (M$_{\odot}$)                          \\  \hline
 
1RXS J040443.5$-$295316   &     1848     &  40.91    &    6.2 \\
 2MASXJ01115114$-$4045426 & 1741   &    41.31    &   6.4   \\
 2MASX J01413249$-$1528016 &  1518 &    41.4   &    6.3    \\
 2MASXJ05014863$-$2253232 & 1510   &    41.32  &     6.3   \\
  2MASX J21124490$-$3730119 &     2248     &  41.04    &   6.5 \\
 2MASX J21531910$-$1514111   &     1852  &     40.65  &     6.1    \\
 2MASXJ21565663$-$1139314 & 1903   &    40.08  &     5.8 \\
 6dF J1117042$-$290233       &     4806     &  41.92    &   7.7 \\ 
 CTSJ13.12        &      1717    &   40.53   &    6.0 \\
 CTSM02.47        &     1964     &  41.26    &   6.5 \\
 EUVEJ0349$-$537    &     3065     &  42.03    &    7.3 \\
 FAIRALL0107      &     1709     &  41.56    &   6.5 \\
 HE1107$+$0129      &     2166     &  41.95    &   6.9 \\
 HE1438$-$0159      &     1673     &  41.26    &   6.3 \\
 IGRJ16185$-$5928   &     4966     &   41.3    &    7.3 \\
 IRAS04576$+$0912   &     1685     &  41.18    &   6.3 \\
 IRAS16355$-$2049   &     3483     &   41.6    &   7.2 \\
 MCG$-$04.24.017    &     1835     &  40.55    &   6.0 \\
 MCG$-$05.01.013    &     1939     &  41.35    &   6.5 \\
 NPM1G$-$17.0312    &     1243     &  39.86    &   5.3 \\ 
 
 RBS1665          &     1725     &  41.36    &   6.4 \\
 RX J0024.7$+$0820   &     1504     &  40.98    &   6.1 \\
 RX J0323.2$-$4931  &     2752     &  41.83    &   7.1 \\
 
 RX J0902.5$-$0700  &     2770     &  41.66    &   7.0 \\

 RX J2301.8$-$5508   &     3228     &  42.25    &   7.5 \\ 
  SDSS J134524.69$-$025939.8 &     1925    &   41.69    &   6.7 \\
 SDSS J144052.60$-$023506.2  &     987.4   &    40.45   &    5.4 \\  
 SDSS J151024.92$+$005843.9 &      2522   &    41.16   &    6.7 \\
 SDSS J153001.82$-$020415.1 & 1699 &      40.49   &    6.0 \\
 SDSS J153705.95$+$005522.8 &   1774   &    41.59    &   6.6 \\
 SDSS J225452.22$+$004631.4 &   2617   &    41.47    &   6.9 \\
 V961349$-$439         &  2906    &   41.26     &  6.8 \\
 WPV85007            &  1582    &    40.92    &   6.1 \\
 Zw037.022          & -----       & -----    & ----- \\ 
 Zw049.106         &  2060      &    40.85    &  6.3 \\
 Zw374.029         &  1276      &    39.82   &  5.4 \\   

\hline
\end{tabular}

\label{tabla.masas}
\end{center}
\end{table*}

\subsection{ The $M_{BH} - \sigma_{\star}$ relation}
\label{relation}
  
Possible correlations between BH masses and properties of the
host galaxies are of fundamental importance to understand the galaxy
formation processes and evolution. The relations between black hole
masses and the velocity dispersion of the bulge have been studied
by several authors, such as \cite{2000ApJ...539L...9F} and \cite{2000ApJ...539L..13G}. \cite{2002ApJ...574..740T}, in the attempt to
relate some galaxy properties to central BHs, found that BH masses for normal galaxies could be estimated with the relation,

\begin{equation}
log \left(\frac{M_{BH}}{ M_{\odot}}\right) = (8.13 \pm 0.06) + (4.02 \pm 0.32) \ log \left(\frac{\sigma_{\star}}{200 \ km \ s^{-1}}\right)
\label{eq:tremaine}
.\end{equation}

\

\noindent On the other hand, in a sample of 14 Seyfert 1 galaxies, \cite{2004ApJ...615..652N} measured the bulge stellar velocity dispersion and determined their SMBH masses using the reverberation mapping technique and found that the Seyfert galaxies followed the same $M_{BH} - \sigma_{\star}$ relation as nonactive galaxies.\\

Usually it is difficult to determine the stellar velocity dispersions in AGNs because of the
strong nonstellar continuum that dilutes the absorption lines of the
underlying stellar populations. On the other hand, gas emission lines are an important feature in
AGNs and they are detectable even in distant galaxies, unlike stellar
absorption lines. Because of this, some authors
(e.g., \cite{2000ApJ...544L..91N}; \cite{2007ApJ...667L..33K}) use, for
instance, [OIII] or [SII] lines instead the stellar velocity dispersion
to study the $M_{BH} - \sigma_{\star}$ relation. For example, when the
width of the narrow [OIII]$\lambda$5007 emission line is used, most
authors found that in general NLS1 galaxies do not follow the $M_{BH}
- \sigma_{\star}$ relation ( \cite{2004ApJ...606L..41G};
\cite{2006MNRAS.367..860B}; \cite{2007AJ....133.2435W}). Nevertheless, \cite{2001A&A...377...52W, 2002ApJ...565..762W}
found that NLS1 follow the relation but with a larger scatter than for normal galaxies.

\begin{figure*}
 \begin{minipage}{\linewidth}
 \begin{center}
  \includegraphics[trim = 3mm 12mm 3mm 130mm, clip, width=1\textwidth]{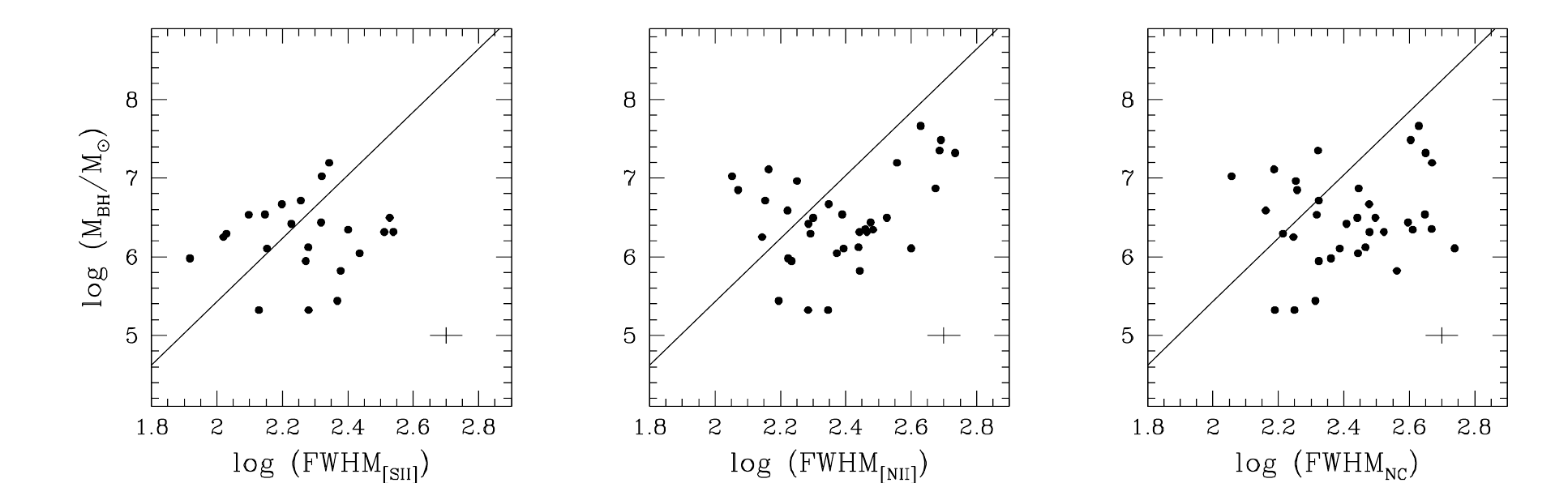}
\end{center}
 \caption{$M_{BH} - \sigma_{\star}$ relation for the galaxies of the sample, using the FWHM of [SII] (left), [NII] (middle), and the narrow component NC of H$\alpha$ (right). All FWHM are in km  s$^{-1}$ and were corrected by the instrumental width. The solid line represents the relation given by Tremaine et al. using FWHM $=$ 2.35 $\sigma_{\star}$. Typical error bars are on each plot.}
\label{fig:tremaine}
 \end{minipage}
\end{figure*}

 In Figure \ref{fig:tremaine} we plot BH masses
against FWHM$_{gas}$, where the solid line represents $M_{BH} -
\sigma_{\star}$ relation for normal galaxies
\citep{2002ApJ...574..740T}. We explore the $M_{BH} - \sigma_{\star}$
relation using FWHM$_{gas}$ instead of $\sigma_{\star}$, where
FWHM$_{gas}$ refers to the FWHM of the emission lines NC of H${\alpha}$,
[NII], and [SII]. We take these widths as
possible proxys of $\sigma_{\star}$, by taking into account the
relation FWHM $=$ 2.3548 $\times$ $\sigma$ for a Gaussian profile of
these lines. 

Taking the uncertainties of 
our determinations on FWHM$_{gas}$ and black hole masses into account, Fig. \ref{fig:tremaine} shows that, in general,
most of the objects lie systematically below the $M_{BH} -
\sigma_{\star}$ relation for normal nuclei. In the case of [NII], around 70 \% of the
galaxies lie below the Tremaine line. The higher deviation is seen
using the NC of H$\alpha$, where $\sim$ 80
\% of the objects lie below the $M_{BH} - \sigma_{\star}$
relation. In the case of [SII] lines, NLS1s are closer to the relation and $\sim$
45 \% of the targets lie above the relation. This agrees with previous
results of \cite{2007ApJ...667L..33K}. Despite the objects being closer to the Tremaine relation, they are slightly below it and we do not find any evidence of correlation among the FWHM of [SII] lines and the BH mass. The cases studied here of [SII], [NII], and NC lines are in agreement with the idea of that NLS1 may mainly reside in galaxies with pseudobulges \citep{2012ApJ...754..146M}. In this scenario, NLS1 do not follow the $M_{BH} - \sigma_{\star}$ because their bulges are intrinsically different from those of other galaxies.\\
Otherwise, the fact that, in general, NLS1 lie systematically below the $M_{BH} -
\sigma_{\star}$ relation for the three studied cases, could imply that NLS1 galaxies have lower BH masses compared to those that follow the relation. It is well known that NLS1 have high Eddington ratios \citep{2004ApJ...608..136W} and related to that, \cite{2000MNRAS.314L..17M} proposed that NLS1 are analogous objects to high-redshift quasars (z $>$ 4). This way, NLS1 may be in an early evolutionary phase that occupies young host galaxies.
It is not very clear how the displacement of NLS1 would be  along the $M_{BH} - \sigma_{\star}$ relation, but the fact that NLS1 have high Eddington ratios suggest that their black hole masses must be rapidly growing  \citep{2005A&A...432..463M, 2005ApJ...633..688M}. In this scenario, the NLS1 tracks on the $M_{BH} - \sigma_{\star}$ would be upward.

\subsection{The intermediate component of H$\alpha$}
\label{intermediate}

In section \ref{measurements} we mentioned that, besides the narrow
and broad components for H$\alpha$, for some objects it was necessary
to include an additional kinematical intermediate component to fit the
line. The sum of only two
components (BC and NC) did not fit the profile of H$\alpha$ well for some galaxies of our sample, thus
giving residuals that were considerably higher than the spectral noise and
requiring an additional component. Indeed, several components for the Balmer emission lines were detected in
some AGNs (\cite{2004A&A...423..909P}, \cite{2009ApJ...700.1173Z}) and NLS1s (\cite{2009arXiv0901.2167Z}, \cite{2008MNRAS.385...53M}). While by definition BC and NC have a direct association with the BLR and NLR,
respectively, described in the standard scenario for Seyferts, the IC
remains clearly related to a well-known physical region. Various
authors, for example, \cite{2009arXiv0901.2167Z} and \cite{2009ApJ...700.1173Z}, claimed that the
emission from the BLR can be described by two Gaussian profiles, one
very broad and an intermediate component. Both of these emissions
could be originated in two different regions, each one with different
velocity dispersions and spatially associated, but there is no clear
scheme yet. To explore this kinematical component, we used our data
from the IC detected in our sample.

\begin{figure*}
 \begin{minipage}{\linewidth}
 \begin{center}
  \includegraphics[trim = 00mm 5mm 0mm 10mm, clip, width=1\textwidth]{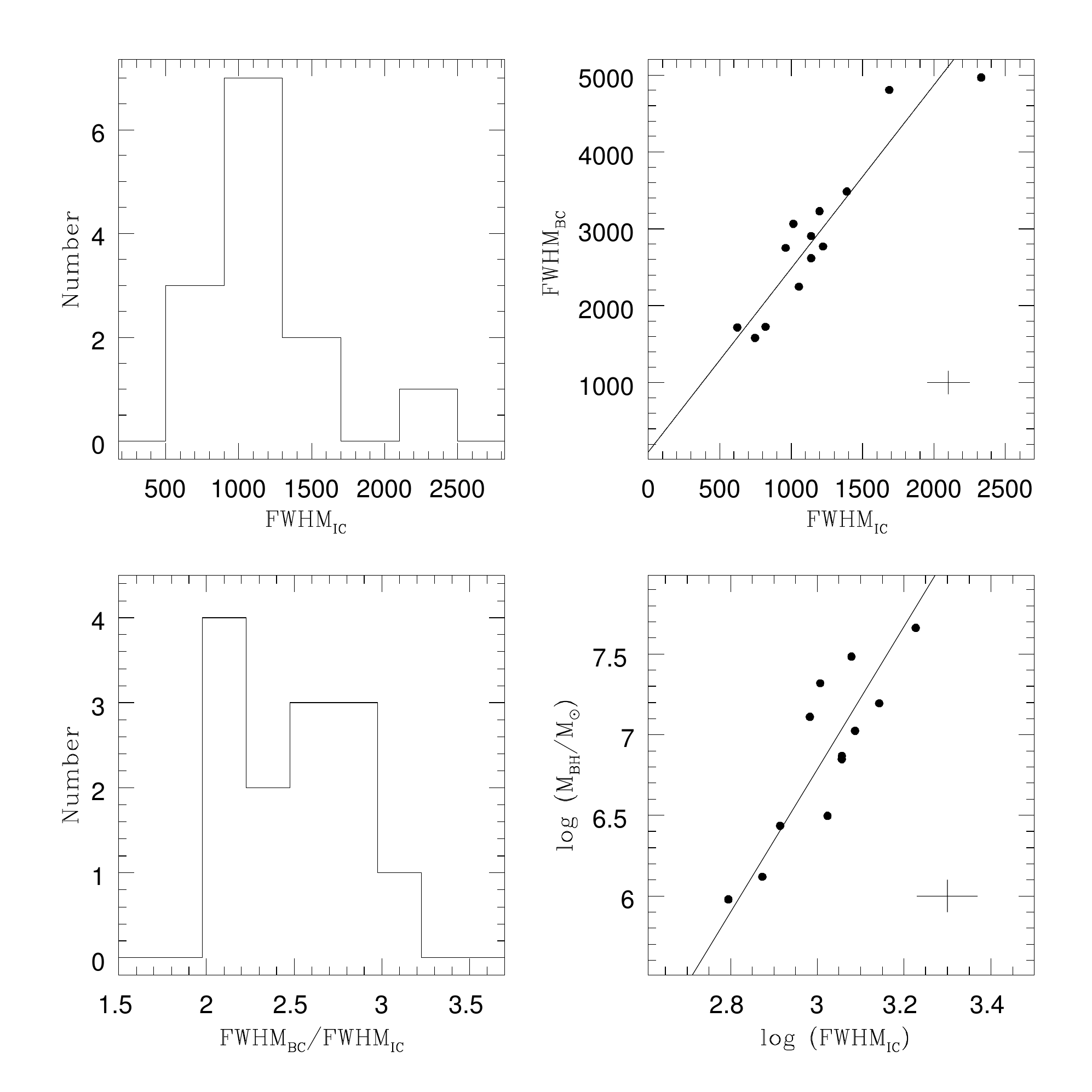}
  \end{center}
  \caption{Histogram of the FWHM of intermediate component (IC) of H$\alpha$ (top left), the relation between the FWHM of BC and the FWHM of IC (top right), the histogram of FWHM$_{BC}$/ FWHM$_{IC}$( bottom left), and the relation between BH masses and the FWHM of IC (bottom right). All FWHM are in units of (km s$^{-1} $). Typical error bars are shown. Solid lines represent the best fits for our data. } 
 \label{fig:HaiFWHM}
 \end{minipage}
\end{figure*}

Following the procedure described
in section \ref{measurements} we determined the IC for 14 galaxies (37
$\%$ of the sample). The main results involving FWHM of
H$\alpha$ IC are presented in Figure \ref{fig:HaiFWHM}.
 The distribution of the FWHM (top left panel of Figure \ref{fig:HaiFWHM}) of IC shows that for 11 out of 14 galaxies, the FWHM of IC is in the range of 600 $-$ 1500 km  s$^{-1}$ with mean value of $\sim$ 1100 km s$^{-1}$. We note that the shape of the distributions for IC, BC, and NC are apparently similar. So, we have applied a Kolmogorov - Smirnov (K-S) test and confirmed that the distribution of the IC is significantly different from those of NC and BC with $P<10^{-8}$ and $P<10^{-5}$, respectively, as being drawn from the same parent distribution. 
This suggests the possibility of three kinematically distinct emitting regions: the well-known BLR and NLR, and an intermediate line region (ILR).

Comparing the FWHM$_{IC}$ and FWHM$_{BC}$ we found that they are strongly related (top right panel). An OLS bisector fit gives
a slope of $\sim$ 2.4 with a small ($<$100 km s$^{-1}$) zero point and a remarkable Pearson correlation coefficient of r$_{p} =$0.93.
We calculated the ratio between the two quantities (bottom left panel), and we find a
mean value of FWHM$_{BC}$/FWHM$_{IC}$ $\sim$ 2.5, which is somewhat lower than
that obtained by \cite{2010SCPMA..53.2307M} of $\sim$3 for H$\beta$. 

We have to keep in mind that the BH masses determinations (eq \ref{eq:GH}) involves the BC, so
we would expect a relation between the FWHM$_{IC}$ and the BH mass, which is actually seen in the bottom
right panel of Figure \ref{fig:HaiFWHM}. An OLS bisector fit
gives a slope of $\sim$ 4.4 and a zero point of log (M$_{BH}$/M$_{\odot}$) $\sim$ $-$6.5. A Pearson 
correlation coefficient of r$_{p} =$ 0.86 shows the close relation between them, suggesting that the dynamics of the ILR is somehow
affected by the central BH.
We tested whether some of these
relations depend on another parameter, such as the luminosity of the
lines, but we did not find any dependence on them.
A particular case is presented by the galaxy HE1107$+$0129, which presents an additional component of $\sim 290$ km  s$^{-1}$ (see Section \ref{curious.cases}), and was not taken into account for the analysis in Figure \ref{fig:HaiFWHM} and in the bottom right plot in Figure \ref{LumBC}. Besides this galaxy, IGRJ16185$-$5928 was not taken into account in the bottom right plot of Figure \ref{fig:HaiFWHM} because of its high deviation from the remaining points.

The tight relationship between FWHM$_{BC}$ and FWHM$_{IC}$ makes us
wonder if there is another quantity governing this behavior. We do not
have any information about the possible location of the origin of the
IC. So far, we only know that, in the frame of the standard scenario
for AGNs, BC and NC originate in the BLR and NLR regions, respectively; these two regions are very different in size, extension, geometry, kinematics,
and physical conditions. The IC detected in our sample tells us that there could
be another physical region. Under the assumption
that the square of the FWHM of the gas decreases with the distance to the center, this ILR is
possibly surrounding the BLR and could be located outside, although
near the BLR with a ratio of mean sizes in a proportion given by
the square of the ratio between the two FWHMs. This way the ratio of both would be R$_{ILR}$/R$_{BLR} \propto$ (FWHM$_{BLR}$/FWHM$_{ILR}$)$^{2} \sim$ 4 $-$ 9. 

\section{Luminosities}
\label{lumyfracpl}

\begin{figure*}
 \begin{minipage}{\linewidth}
 \begin{center}
 \includegraphics[trim = 00mm 5mm 0mm 10mm, clip, width=1\textwidth]{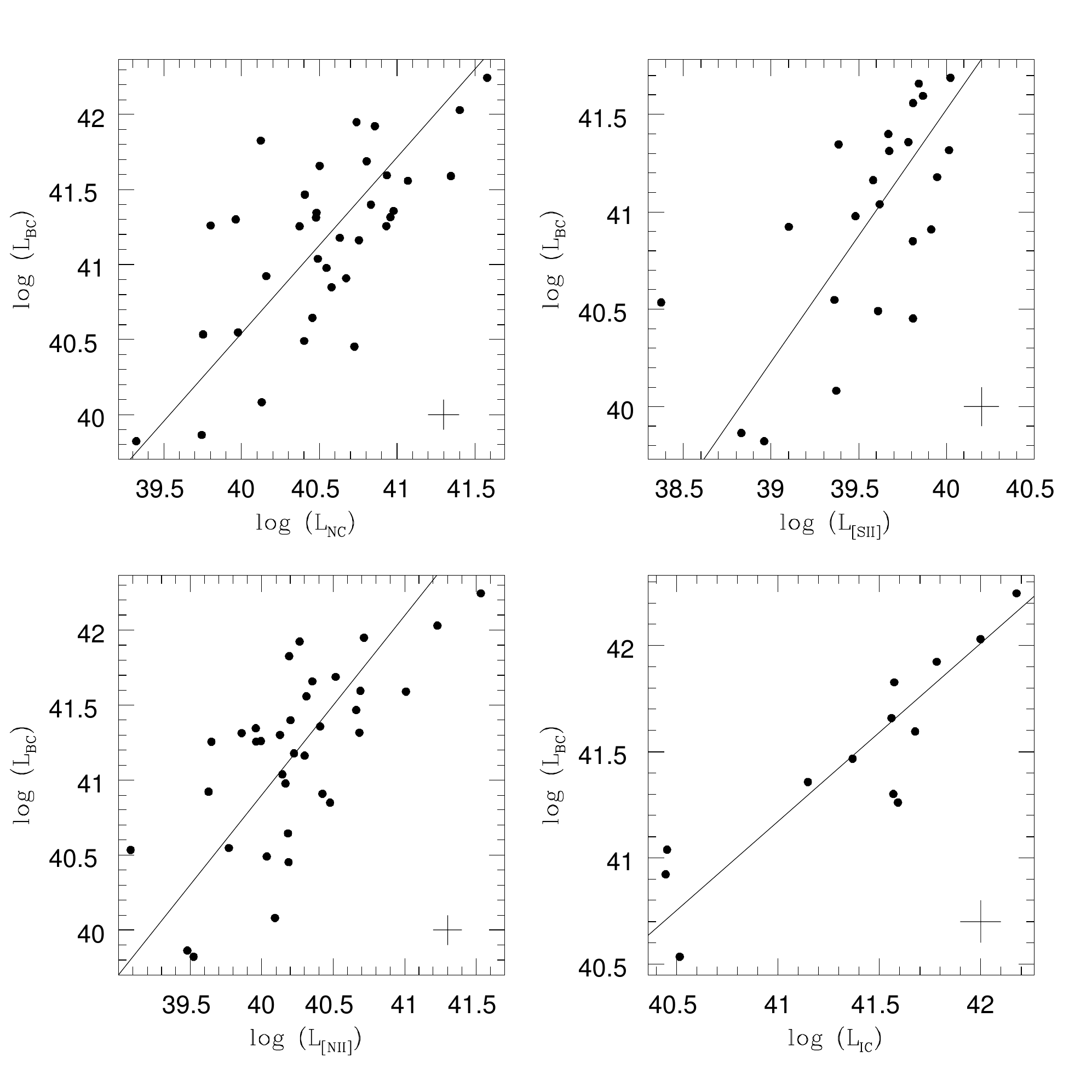}
  \end{center}
 \caption{Luminosity of BC compared with those of the NLR: NC (top left), [SII] (top right), [NII] (bottom left), and with the IC (bottom right). All luminosities are in units of (ergs s$^{-1}$). Solid lines represent the best fits for our data. Typical error bars are shown in each plot.}
 \label{LumBC}
 \end{minipage}
\end{figure*}

We determined the luminosity for each component of H$\alpha$, as well
as for [NII] and [SII] lines. Figure \ref{LumBC} shows the
luminosities of BC compared to those of NC (top left panel),
[NII]$\lambda$6584 (top right), [SII]$\lambda$6731 (bottom left), and
IC (bottom right).

The luminosities of BC are systematically higher
than those of the lines originated at the NLR ([NII], [SII], and NC), and that of the IC, and a relation is evident between them. 
We found that the luminosity of the BC is typically eight times the luminosity of the NC, and only two objects (2MASXJ21565663$-$1139314 and SDSSJ144052.60$-$023506.2) present similar or
lower luminosity of the BC compared to that of the NC. A similar trend is
observed when comparing BC to [NII]$\lambda$6584 and
[SII]$\lambda$6731. 
In the case of the IC, the
luminosities of the BC are slightly higher for 10 out of 13
objects, and the remaining present weak BCs, and thus they have very
low luminosities.

We applied OLS bisector fits to these data, which give slopes of around 1.2, 1.2, and 1.3 for NC, [NII], and [SII] lines, respectively. The Pearson coefficients are typically around r$_{p} =$ 0.67 for these three relations. The slopes of the relations between the BC luminosity and that of the lines arising in the NLR are similar, which could suggest that the ionizing source is the same in the three cases.
Interestingly, there is a stronger
relationship between the BC and IC luminosities: the OLS bisector fit gives a slope of $\sim$ 0.84 and a Pearson
correlation coefficient of 0.90, which probably implies a dependence among their
emitting regions and also possibly on their physical conditions. The fact that for this case the slope is slower than that for narrow lines could suggest a possible physical difference between the emitting regions (ILR and NLR). Curiously, comparing L$_{BC}$ to L$_{IC}$, they are quite similar with a mean value of L$_{BC}$/L$_{IC}\sim$1.5. This interesting result tells us that the luminosities of both regions are indeed comparable. The geometry of the emitting regions and the presence of different amount of gas in each region probably play an important role on their luminosities. Nonetheless, this is not the case for the ILR, for which, even though physical 
conditions are not well known yet, the relation between BC and IC is 
stronger than for NLR lines.

\section{Some curious cases}
\label{curious.cases}

Two objects show higher velocity dispersion for the IC. These galaxies are 6dF J1117042$-$290233 and IGRJ16185$-$5928, which have
FWHM of $\sim$ 1700 and $\sim$ 2000 km s$^{-1}$, respectively. They also show the higher velocity dispersion for the BC, $\sim$ 5000 km s$^{-1}$, and the highest black hole masses, log (M$_{BH}$/M$_{\odot}$) $>$ 7.3.

We do not detect any BC for Zw037.022, which only has a NC for this Balmer line of $\sim$ 130 km s$^{-1}$. 
According to \cite{1996ApJS..106..341M} and \cite{2008A&A...484..897K}, this object has been classified as Sy1 and NLS1, respectively.
On the other hand, according to the spectral classification of the Sloan Digital Sky Survey Data Release 12 \citep[SDSS DR12;][]{2015ApJS..219...12A, 2011AJ....142...72E}, Zw037.022 is a star forming galaxy. This classification is based on whether the galaxy has detectable emission lines that are consistent with star formation according to the criteria log ([OIII]$\lambda$5007/H$\alpha$) $<$ 0.7 $-$ 1.2(log ([NII]$\lambda$6584/H$\alpha$) $+$ 0.4).
In our case, we find that log ([NII]$\lambda$6584/H$\alpha$)$= -$0.8, log ([SII]$\lambda$6716/H$\alpha$)$= -$0.8, and  log ([SII]$\lambda$6731/H$\alpha$)$= -$0.9. These results agree with the idea of that this object must be classified as a star forming galaxy. In addition, the fact that this galaxy does not show a BC supports the idea that it is neither a Sy1 nor a NLS1 galaxy.
 
As stated in Section \ref{intermediate}, the object HE1107$+$0129 shows an additional component of $\sim 290$ km  s$^{-1}$. We tried to fit the spectrum of this galaxy with only two components for H$\alpha$, but the residuals were much higher than the noise of the spectrum, and so another component was necessary. This extra component is comparable with the FWHM distribution of NC and shows a blueshift of $\sim$ 360 km s$^{-1}$ compared to the NC (FWHM of $\sim$ 190 km s$^{-1}$). We analyzed the relation between the FWHM of BC and NC for the sample and the medium value is FWHM$_{BC}$/FWHM$_{NC}$ $\sim$ 8. For this galaxy, the ratio between the FWHM$_{BC}$ and the FWHM of this additional component is also $\sim$ 8. These facts suggest that this extra component is emitted from the NLR and not from an ILR. For this reason, it was not taken into account in the 
analysis of Section \ref{intermediate}.

\section{Final remarks}
\label{final}

 We have observed and analyzed the spectroscopic data of a sample of 36
NLS1 galaxies, 27 of which are in the Southern Hemisphere. We performed careful Gaussian decomposition to the main emission lines in the red spectral range. Several components were fitted to the global profile for H$\alpha+$[NII]$\lambda\lambda$6548,6583. This allowed us to estimate the virial BH masses of the galaxies. The obtained values for our sample are in the
range log (M$_{BH}$/M$_{\odot}$) $= 5.3 - 7.7$, where the mean value is around log (M$_{BH}$/M$_{\odot}$) $=$
6.2. The black hole masses are in the range
log (M$_{BH}$/M$_{\odot}$) $= 5.7 - 7.3 $ for 32 out of 36 galaxies. These values are lower that those found in broad line Seyfert 1 galaxies, confirming that on average, NLS1 have lower BH masses \citep[e.g.,][]{2007ApJ...667L..33K,2004AJ....127.1799G}.

  We tested the narrow component of H$\alpha$, [NII], and [SII] lines as proxys of $\sigma_{\star}$, using the FWHM of these emission lines to examine the $M_{BH} - \sigma_{\star}$ relation for normal galaxies. Taking into account the uncertainties of our determinations on the FWHM of the emission lines and black hole masses, we found that in general most of our NLS1s lie below the $M_{BH} - \sigma_{\star}$ relation for normal galaxies. The fact that NLS1 have lower black hole masses, and taking into account that they have high acretion rates \citep{2004ApJ...608..136W} and therefore their black hole mass are growing quickly  \citep{2005A&A...432..463M, 2005ApJ...633..688M}, could suggest that such AGNs are at an early stage of nuclear activity. Although in the case of the [SII] lines the galaxies seem to be closer to the relation than the other lines (in agreement with \cite{2007ApJ...667L..33K} results), we do not see any correlation between the FWHM of [SII] lines and the BH mass. The fact that the 
three tested emission lines put the NLS1 below the $M_{BH} - \sigma_{\star}$ relation could be in agreement with \cite{2012ApJ...754..146M}, who found that such AGNs are mainly hosted by galaxies with pseudobulges. Related to this, NLS1 would not follow the $M_{BH} - \sigma_{\star}$ because their bulges are different from those of other AGN types.   

 We studied the intermediate component of H$\alpha$ found in 13 out of 36 galaxies. Comparing the distribution of velocities between the three components of H$\alpha$, we see that the FWHM of BC is between 900 $-$ 3200 km s$^{-1}$ with a peak in 1600 km s$^{-1}$  for most objects; the IC is in the velocity range 700 $-$ 1500 km s$^{-1}$ for most galaxies, with most objects showing FWHM of 1100 km  s$^{-1}$; while for the NC the velocity range is 180 $-$ 500 km s$^{-1}$ with a peak in 220 km s$^{-1}$. We note a high correlation between FWHM$_{IC}$ and FWHM$_{BC}$, with a Pearson correlation coefficient of r$_{p} =$ 0.93. We found for our sample that FWHM$_{BC}$ / FWHM$_{IC}$ is in a range of 2.1 $-$ 3 with a mean value of $\sim$ 2.5, which is somewhat lower than that obtained by \cite{2010SCPMA..53.2307M} of $\sim$3 for H$\beta$ . We also performed Kolmogorov-Smirnov (K-S) tests to FWHM$_{BC}$ and FWHM$_{IC}$, and to FWHM$_{NC}$ and FWHM$_{IC}$, which give probabilities of $P<10^{-5}$ and $P<10^{-8}$ as being 
drawn from the same parent distribution, respectively. All these results point out the possibility of three kinematically distinct emitting regions and show that BC and IC are somehow linked, possibly implying that IC could arise in a region surrounding the BLR. Taking into account that BH masses depend on FWHM$_{BC}$, and FWHM$_{IC}$ correlates with FWHM$_{BC}$, it is expected to find a correlation between BH masses and FWHM$_{IC}$. In fact, this correlation has a Pearson correlation coefficient r$_{p} =$0.86, suggesting that the dynamics of this emitting region is clearly affected by the central engine.\\
The presence of the additional intermediate component mainly affects the NC. If only two components were used to fit the H$\alpha$ line, NC would be much broader and would show a greater amount of line flux. In the case of the BC, it does not vary significantly; in general it could increase its FWHM by $\sim$10\% and decrease the flux by $\sim$10$-$25\%. Contrary to what we expected, BH masses in general decrease $\sim$10$-$20\% if one consider only two components for this Balmer line.\\
 We also studied the emission line luminosities of broad and narrow lines.
There are some tendencies between the BC luminosity and those of the narrow lines
[NII], [SII], and NC with Pearson correlation coefficients around 0.67 for the three cases, indicating that the emission in BLR and NLR are proportional. Interestingly, their slopes are similar, around 1.2, as derived from OLS bisector fits. The relation is even stronger for the IC with a slope of around 0.84 and a Pearson correlation coefficient of 0.90. The fact that the slope is different for the NC and the IC lines would indicate possible physical difference between the regions. Interestingly, we found that the luminosities of BC and IC are comparable with a mean value L$_{BC}$/L$_{IC}$ $\sim$1.5. Related to this, the geometry of the emitting regions, despite distinct amounts of gas present in each one, play an important role. Even though the physical conditions of 
the ILR are not well known yet, our results suggest that this region is different from the BLR and NLR, with 
properties that are somehow related and strongly linked to those of the BLR.

\begin{acknowledgements}
     
    This work was partially supported by Consejo de Investigaciones Cient\'ificas y T\'ecnicas (CONICET) and Secretar\'ia de Ciencia y T\'ecnica de la Universidad Nacional de C\'ordoba (SecyT). We want to thank the anonymous referee, whose very useful remarks helped us to improve this paper.
 This research has made use of the NASA/IPAC Extragalactic Database (NED) which is operated by the Jet Propulsion Laboratory, California Institute of Technology, under contract with the National Aeronautics and Space Administration. 
\end{acknowledgements}

%
\bibliographystyle{aa} 
\bibliography{referencias} 
%

\end{document}